\def\preprint{0}                
\def\preprint{1}                
\def\comment#1{}
\if\preprint1
        \documentclass[usenatbib]{mn2e}
        \usepackage{times}
        \usepackage{graphicx}
        \usepackage{calc}

\else
        \documentstyle[astrop-bib,referee,times]{mn}
        \newcommand{\includegraphics}[1]{}
\fi

\def\oversim#1#2{\lower0.5pt\vbox{\baselineskip0pt \lineskip-0.5pt
     \ialign{$\mathsurround0pt #1\hfil##\hfil$\crcr#2\crcr\sim\crcr}}}
\def\gsim{\mathrel{\mathpalette\oversim>}}    
\def\lsim{\mathrel{\mathpalette\oversim<}}    

\title[Carbon stars 
in the Small Magellanic Cloud]{Spitzer  spectroscopy of 
carbon stars in the Small Magellanic Cloud}
\author[]{Eric Lagadec$^{1}$ \thanks{E-mail:
eric.lagadec@manchester.ac.uk},
               Albert~A.~Zijlstra$^1$,
   G.C. Sloan$^2$,    Mikako Matsuura,$^{3,1}$
        Peter Wood$^4$, G.J. Harris$^5$,\newauthor  
Jacco Th. van Loon$^6$, J.A.D.L. Blommaert$^7$, 
S. Hony$^8$, M.A.T. Groenewegen$^7$,    M.W. Feast$^9$,\newauthor 
 P.A. Whitelock$^{9,10,11}$, J.W. Menzies$^9$,
M.-R. Cioni$^{12}$, H. Habing$^{13}$, L.B.F.M. Waters$^{14}$\\
\\
$^1$University of Manchester, School of Physics \&\ Astronomy, 
          P.O. Box 88, Manchester M60 1QD, UK\\
$^2$Department of Astronomy, Cornell University, 108 Space Sciences 
            Building, Ithaca NY 14853-6801, USA\\ 
$^3$  Division of Optical and IR Astronomy, National Astronomical Observatory 
     of Japan, Osawa 2-21-1, Mitaka, Tokyo 181-8588, Japan\\     
$^4$Research School of Astronomy and Astrophysics, 
          Australian National University,  Cotter Road, Weston Creek,
          ACT 2611, Australia\\
$^5$Department of Physics and Astronomy, University College London, 
      Gower Street, London WC1E 6BT, UK\\
$^6$Astrophysics Group, School of Physical \&\ Geographical Sciences, 
Keele University, Staffordshire ST5 5BG, UK\\
$^7$Instituut voor Sterrenkunde, K.U. Leuven,
Celestijnenlaan 200 D, B-3001 Leuven, Belgium\\ 
$^8$CEA, DSM, DAPNIA, Service d'Astrophysique, C.E. Saclay, F-91191 Gif-sur-Yvette Cedex, France\\ 
$^9$ Department of  Astronomy, University of Cape Town, 7701 Rondebosch, 
           South Africa\\
$^{10}$South African Astronomical Observatory, PO Box 9, 7935
Observatory,  South Africa\\
$^{11}$NASSP, Department of Mathematics \&\ Applied Mathematics 
University of Cape Town, 7701, Rondebosch, South Africa\\
$^{12}$Institute for Astronomy, University of Edinburgh, Royal 
Observatory, Blackford Hill, Edinburgh EH9 3HJ, UK\\
$^{13}$Sterrewacht Leiden, Niels Bohrweg 2, 2333 RA Leiden, 
  The Netherlands\\
$^{14}$Astronomical Institute, University of Amsterdam, Kruislaan 403, 
           1098 SJ Amsterdam, The Netherlands\\
}
\pubyear{2006}

\begin{document}

\date{Accepted . Received}

\pagerange{\pageref{firstpage}--\pageref{lastpage}} \pubyear{2002}

\maketitle

\label{firstpage}

\begin{abstract}
We present {\it Spitzer Space telescope} spectroscopic observations of 14 carbon-rich
AGB stars in the Small Magellanic Cloud.  SiC dust is seen in most of the
carbon-rich stars but it is weak compared to LMC stars.  The SiC feature is
strong only for stars with significant dust excess, opposite to what is
observed for Galactic stars. We argue that in the SMC, SiC forms at lower
temperature than graphite dust, whereas the reverse situation occurs in the
Galaxy where SiC condenses at higher temperatures and forms first. Dust input
into the interstellar medium by AGB stars consists mostly of carbonaceous dust,
with little SiC or silicate dust. Only the two coolest stars show a 30-micron
band due to MgS dust. We suggest that this is due to the fact that, in the SMC,
mass-losing AGB stars generally have low circumstellar (dust) optical depth and
therefore effective heating of dust by the central star does not allow
temperatures below the 650 K necessary for MgS to exist as a solid.  Gas phase
C$_2$H$_2$ bands are stronger in the SMC than in the LMC or Galaxy. This is
attributed to an increasing C/O ratio at low metallicity. We present a
colour-colour diagram based on {\it Spitzer} IRAC and MIPS colours to discriminate
between O- and C-rich stars. We show that AGB stars in the SMC become carbon
stars early in the thermal-pulsing AGB evolution, and remain optically visible
for $\sim 6 \times 10^5$ yr. For the LMC, this lifetime is $\sim 3 \times
10^5$ yr. The superwind phase traced with {\it Spitzer} lasts for $\sim 10^4$ yr.
Spitzer spectra of a K supergiant and a compact HII region are also given.
\end{abstract}



\section{Introduction}
The late stages of the evolution of low and intermediate mass stars
(hereinafter LIMS) are characterised by intense mass loss.  This so-called
superwind occurs during the Asymptotic Giant Branch (AGB) phase. The mass loss
leads to the formation of a circumstellar envelope composed of gas and
dust. The chemical composition depends strongly on the C/O abundance ratio.
CO is one of the first molecules to form and is very stable and unreactive. If
C/O$\,>$1 by number, O will be trapped in the CO molecules, leading to a
chemistry dominated by molecules such as C$_2$, C$_2$H$_2$, HCN and SiC dust
grains (carbon stars). On the other hand, if C/O$\,<$1 we will observe
oxygen-rich stars showing SiO, OH, H$_2$O molecules and silicate dust.

The study of this mass loss is of high astrophysical interest, and impacts on
the chemical evolution of galaxies. Indeed, the mass loss from LIMS
contributes to roughly half of all the gas recycled by stars (Maeder
1992). Mass loss from LIMS is one of two main sources (along with WR stars and
supernovae associated with massive stars), of carbon in the universe (Dray et
al. 2003). LIMS are also the main source of heavy s-process elements (e.g. Ba,
Pb) and, when including post-AGB evolution (e.g. novae), the major stellar
source of lithium.

The mass-loss process is, however, not fully understood. The mass loss results
from a complicated interplay between stellar processes (turbulent convection,
pulsation) and circumstellar processes (pulsation-driven shocks, radiation
pressure) where especially the role of the dust composition is disputed
(Woitke 2006).  The effect of metallicity on the mass-loss rates is poorly
understood and may vary with dust chemistry. Observational studies show no
difference between peak gas mass-loss rates for LMC and SMC stars (e.g.  van
Loon 2006). Theoretical studies (e.g. Bowen \& Willson 1991) predict that the
mass-loss rates in AGB stars depend on the metallicity: a lower metallicity
leads to a lower dust-to-gas ratio and less efficient dust-radiation pressure.
The effect may differ between mineral (e.g. silicates) and non-mineral dust.
In O-rich stars, all dust species are minerals (containing Si, Al for
respectively silicates and corondum), whilst in C-rich stars dust contains
both mineral (e.g SiC) and non-mineral components (e.g. soot, amorphous
carbon).

The sensitivity reached by the {\textit Spitzer Space Telescope} (Werner et
al. 2004) enables, for the first time, the determination of mass-loss rates
from stars of different masses all along the AGB sequence at the distance of
the Magellanic Clouds. The distances to these two galaxies are also relatively
well-known and the metal abundance of stars within can be estimated using
age-metallicity relations. This allows one to measure absolute (dust)
mass-loss rates for stars with known bolometric magnitudes and metallicities.

We have therefore conducted a survey of mass-losing stars in the Small and
Large Magellanic Clouds (hereinafter SMC and LMC respectively). The aim of
this project is to empirically calibrate the mass-loss rate of AGB stars as a
function of mass, luminosity and metallicity.  Here we present the data from
the SMC. The data from the LMC are presented in Zijlstra et al. (2006). A
study of the mass-loss rates will be presented in forthcoming papers.

\begin{figure}
\includegraphics[width=\columnwidth,clip=true]{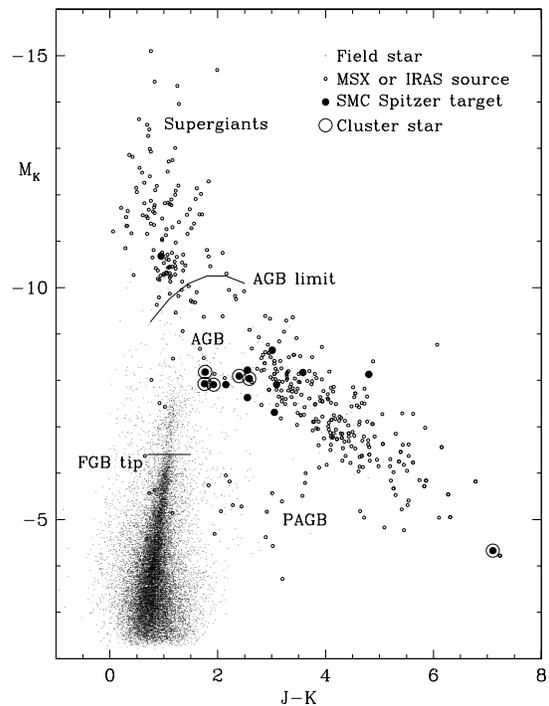}
\caption{\label{colours.eps} 
The M$_{\rm K}$, J$-$K diagram for the observed sample of mid-IR sources in
the SMC (large filled circles).  Stars in the cluster NGC 419 are circled.
For orientation, we also show a large sample of mid-infrared sources in the
LMC from MSX (small open circles) and field stars from a small area of the LMC
bar (small dots) - see also Zijlstra et al. (2006).  AGB stars are confined
approximately to the region below the line marked ``AGB limit". Stars above
this limit are red supergiants with masses above $\sim$8 M$_{\odot}$, or
perhaps foreground stars.  Distance moduli of 18.54 and 18.93 have been
assumed for the LMC and SMC, respectively (Keller \&\ Wood 2006).  }
\end{figure}

\section{Target selection}

We selected 17 stars in the SMC to obtain a sample of stars of lower
metallicity than the LMC stars described in Zijlstra et al. (2006).  The stars
were all AGB stars, with preference being given to stars in the populous
intermediate-age cluster NGC 419, stars with past infrared monitoring showing
large-amplitude variability or stars with some ISO detection (Cioni et
al. 2003).

The observed stars are shown in the M$_{\rm K}$ vs J$-$K diagram in
Fig~\ref{colours.eps}.  Fourteen of the intended targets were observed with
{\it Spitzer}, while peak-up failures occurred for 3 objects.  In one
case, the peak-up occurred on a noise spike so no target was observed, while
in the other two cases a brighter object in the peak-up image led to the
detection of an HII region (MB 88, from Murphy \&\ Bessell 2000) and a red
supergiant of spectral type K (PMMR 52, from Pr\'evot et al. 1983).  The latter
object is the single object lying in the supergiant region of Figure
\ref{colours.eps}.

The observed stars in the cluster NGC 419 whose names contain LE (NGC 419 LE
16, NGC 419 LE 18, NGC 419 LE 27 and NGC 419 LE 35) were discovered by Lloyd
Evans (1980) using photographic surveys in the V and I bands. The two other
stars from this cluster, NGC 419 IR1 and NGC 419 MIR1 were discovered during a
survey with ISOCAM (Tanab\'{e} et al. 1997). LEGC 105 was discovered during an
optical survey of the SMC (Lloyd Evans et al. 1988).  RAW 960 appeared in the
Rebeirot et al. catalogue (1993). IRAS\,00554$-$7351 is described in Whitelock
et al. (1989). Stars with names beginning with ISO were selected from the
ISO/MACHO catalogue of variable AGB stars in the SMC (Cioni et al. 2003). The
variable star GM\,780 was discovered by Glen Moore using UK Schmidt plates but
his work is unpublished. Literature data is summarised in Table
\ref{lit.dat}. For LEGC\,105, Lloyd Evans et al. (1988) suspect that the
pulsation is irregular but their approximate period is confirmed with MACHO.
RAW\,960 has a very short period and is clearly a semi-regular rather than a
Mira variable.

All 14 AGB stars that were observed turned out to be carbon stars.  The
dominance of C-stars is at variance with the MSX-based classification scheme
of Egan et al. (2001) according to which the sample stars would be a mixture
of carbon-rich and oxygen-rich stars.  Our LMC sample of AGB stars also showed
that the Egan et al. classification scheme is ineffective at separating M- and
C-stars on the AGB (Zijlstra et al. 2006).

The AGB stars in our SMC sample have JHK colours which are reddened compared
to stellar colours, but mostly not extremely so. The J$-$K versus M$_{\rm K}$
diagram is shown in Fig. 1. The SMC sample has predominantly lower
circumstellar extinction (bluer J$-$K colours) than the LMC sample in Zijlstra
et al. (2006). The cluster stars from NGC 419 are mostly at the blue end of
the AGB sequence, but they also include the reddest star in the sample (NGC
419 MIR1).  IRAS 00554$-$7351 is also very red with J$-$K$ = 5$.

\begin{figure*}
\includegraphics[width=\textwidth,clip=true]{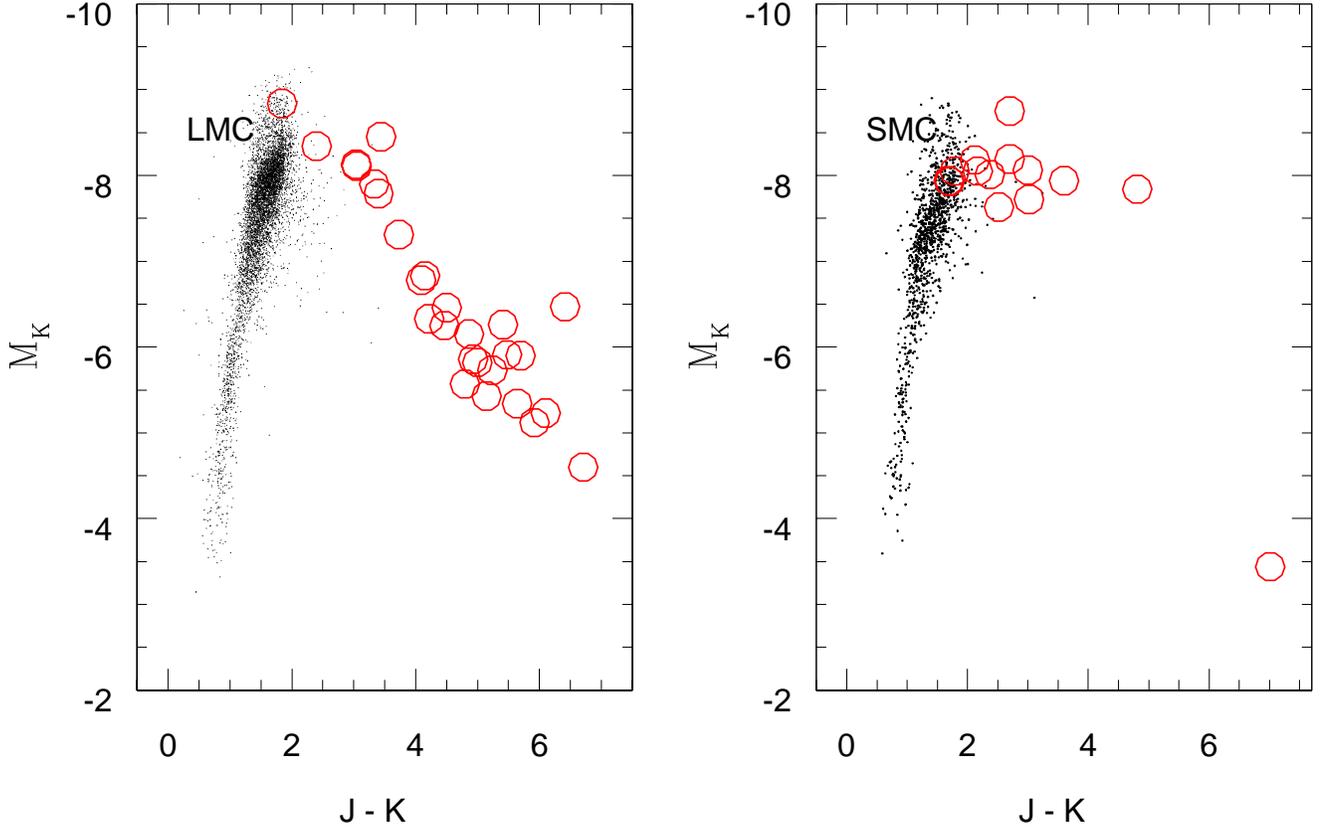}
\caption{\label{allstars.ps} 
The M$_{\rm K}$, J$-$K diagram for all known optical carbon stars in the
LMC and SMC. Samples are taken from Kontizas et al. (2001) and Rebeirot et
al.  (1993); infrared magnitudes are from 2MASS.  Open circles show the
Spitzer targets discussed in Zijlstra et al. (2006) and in this paper.
LMC photometry for these is from Groenewegen et al. (2007): where no J is
available, we assumed $\rm J-K\sim 0.6+1.84\times (H-K)$. The J$-$K$\approx 7$
for NGC418\,MIR1 was estimated from its K$-$L colour.
 }
\end{figure*}

Fig. \ref{allstars.ps} shows the colour-magnitude diagrams for known optical
carbon stars in the Clouds. Open circles show the {\it Spitzer} samples of Zijlstra
et al. (2006) and this paper.  The distribution shows that we are observing
the peak of the luminosity range of the optical carbon star distribution.  The
LMC sample contains a larger fraction of redder objects with higher dust
mass-loss rates than the SMC sample. The gas mass-loss rates may be more
compatible (e.g. Matsuura et al. 2006, van Loon 2006). However, this point
should be taken into account when comparing the samples, as later in this
paper.

\section{Observations}
\subsection{Spitzer}

The observations were made with the InfraRed Spectrograph (IRS, Houck et al.
2004), on board the \textit{Spitzer Space Telescope}. We used the Short-Low
(SL) and Long-Low (LL) modules to cover the wavelength range 5-38$\mu$m.  The
SL and LL modules are each divided in two spectral segments, together known as
SL2, SL1 , LL2 and LL1; a ``bonus'' order covering the overlap between the two
modules is also available.  The data reduction is similar to that described in
Zijlstra et al. (2006).  The raw spectra were processed through the {\it Spitzer}
pipeline S12. We replaced the bad
pixels by values estimated from neighbouring pixels. The sky subtraction was
done by differencing images aperture by aperture in SL and nod by nod in
LL. We used the software Spice ({\it Spitzer} IRS Custom Extractor) to extract the
spectra. The flux calibration was made using the reference stars HR 6348 (K0
III) in SL and HR 6348, HD 166780 (K4 III) and HD173511 (K5 III) in LL. The
spectra were individually extracted from the individual images. Both nods in
both apertures were then joined simultaneously, recalculating the errors in
the process by comparing the nods. The different nods were averaged, using the
differences to estimate the errors. The different spectral segments were
combined using scalar multiplication to eliminate the discontinuities due to
flux lost because of pointing errors. The different segments were also trimmed
to remove dubious data at their edges. We also retained the bonus order where
it was valid.  We obtained a standard wavelength calibration accuracy of
0.06$\mu$m in SL and 0.15$\mu$m in LL.  The calibration process is detailed in
Sloan et al. (2003).

Even after sky removal, some interstellar emission lines remain on the
spectra, mostly observed longward of 30$\mu$m.  

Two of the observed objects in NGC 419, LE 35 and LE 27, showed large
discrepancies between SL and LL.  The apertures of these two modules are
perpendicular to each other on the sky. An acquisition error may have led to
the star being missed in one aperture but not the other, or confusion in the
cluster may have caused spectrum extraction problems. The SL spectra are those
of AGB stars, with flux densities consistent with the MSX flux. The LL spectra
are too faint and featureless. We therefore consider the SL observations to
have been successful and will only discuss their SL spectra.  However,
mispointings can affect the wavelength calibration, flux calibration and even
the slope of the spectral energy distribution (SED). We can therefore not be
as confident of these spectra as we are for the remaining stars.

\begin{figure*}
\begin{center}
\includegraphics[width=13.5cm]{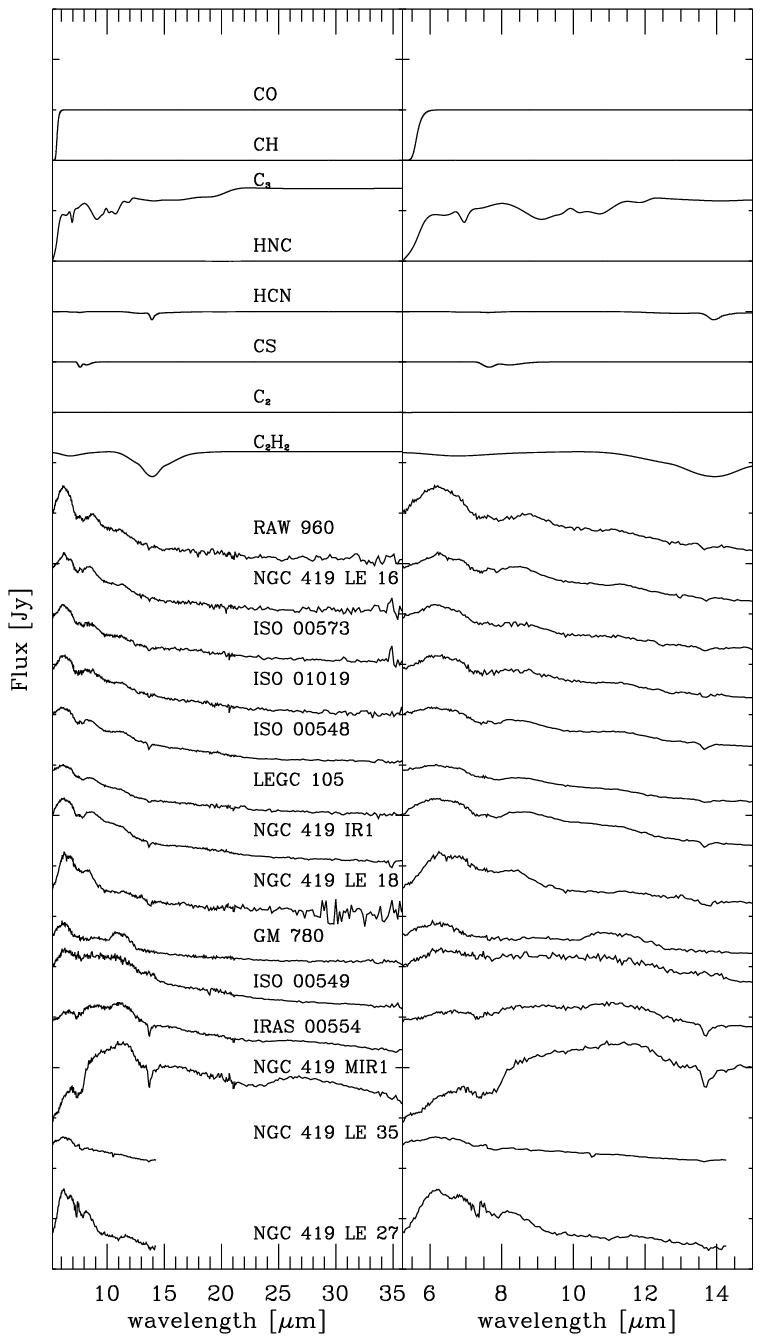}
\caption{\label{spec_multi_offset_bands.ps}Spectra of the carbon-rich 
stars observed in the SMC. The two stars at the bottom have been observed only
in SL. All other stars are ordered by dust temperature. Blue stars are at the
top and red ones at the bottom. Molecular templates are shown at the top
(see Sect. 4.3).}
\end{center}
\end{figure*}

The spectra of the observed AGB stars, ordered by dust temperatures (see
Sect. 6), are presented in Fig.~\ref{spec_multi_offset_bands.ps}.  The
molecular bands discussed below show that all 14 objects are carbon-rich
stars.  We calculated equivalent magnitudes of our sources for IRAC 8$\mu$m
and MIPS 24$\mu$m, by convolving these filter band passes with our
spectra. These are listed in Table
\ref{targets.dat}.

\begin{figure}
\includegraphics[width=9cm,height=6cm]{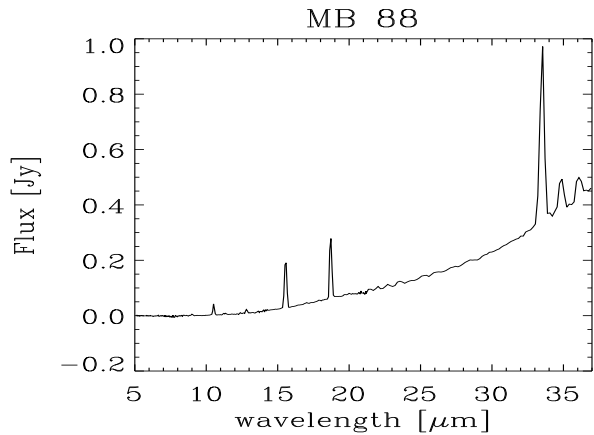}
\caption{\label{iso518}Spectrum of the red object MB\,88. This object shows 
emission lines and could be associated with an H{\sc ii} region. Emission
lines from S{\sc iv} (10.51$\mu$m), Ne{\sc ii} (12.81$\mu$m), S{\sc iii} 
(18.71 and 33.48$\mu$m), Si{\sc ii} (34.83$\mu$m) and Ne{\sc iii}
 (35.97$\mu$m) are observed.
}
\end{figure}

\begin{figure}
\includegraphics[width=9cm,height=6cm]{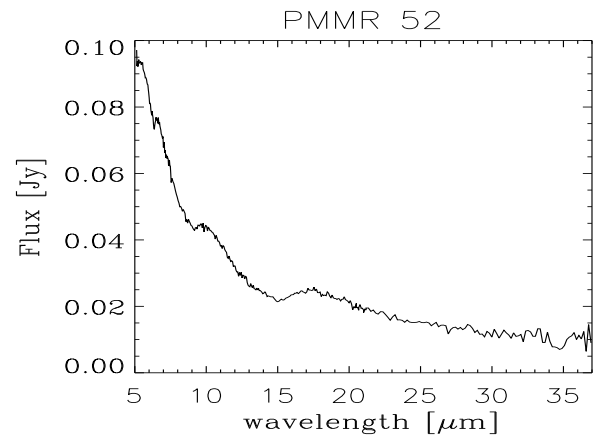}
\caption{\label{gm106.a.lo.ap.ps}Spectrum of the oxygen-rich star PMMR\,52. 
Typical spectral features of silicates are observed at 9.8$\mu$m and 18$\mu$m. }
\end{figure}

Finally, we present the spectra of the two objects acquired accidently
following peakup failure.  MB\,88 (Fig. \ref{iso518}) is a red object with
emission lines, a spatial diameter of about 8 arcsec, and with several
embedded stars.  It is almost certainly an HII region.  Emission lines from
S{\sc iv} (10.51$\mu$m), Ne{\sc ii} (12.81$\mu$m), S{\sc iii} (18.71 and
33.48$\mu$m), Si{\sc ii} (34.83$\mu$m) and Ne{\sc iii} (35.97$\mu$m) are
observed.  Its spectrum is not analysed further here.  The spectrum of the red
supergiant PMMR 52 can be found in Fig. \ref{gm106.a.lo.ap.ps}. It is
discussed briefly in Sect. \ref{sectpmmr52}.

\subsection{Bolometric magnitudes and initial masses}

 We determined the bolometric magnitudes of the observed stars assuming a SMC
distance modulus of $18.93\pm0.024$ (Keller \& Wood 2006) ($\sim$60\,kpc).
JHKL photometry was taken nearly simultanously with the {\it Spitzer}
observations. These observations were made using the 2.3m Telescope at Siding
Spring Observatory (SSO, Australia). The filters used were centred at
1.28$\mu$m (J), 1.68$\mu$m (H), 2.22$\mu$m (K) and 3.59$\mu$m (L). The
observations are described in Groenewegen et al. (2007). Zero flux was assumed
at the frequencies 0 and $3 \times 10^{16}$ Hz to estimate the bolometric
magnitudes.  The flux from the star was then estimated by integrating under
the resulting SED.  The line-of-sight extinction towards the SMC of $E(\rm
B-V)=0.12$ (Keller \&\ Wood 2006) was ignored.  The bolometric
magnitudes are presented in Table \ref{targets.dat}.  

Simultaneous observations in the K-band are important to obtain a reliable
instantaneous luminosity estimate. However, the magnitudes are single-epoch
measurements and therefore subject to pulsation-induced variability.  Earlier
(literature) values are listed in Table 1. These generally agree to within 0.5
mag of the current measurements, consistent with the variability expected for
these stars.

\begin{figure}
\includegraphics[width=8.5cm,clip=true]{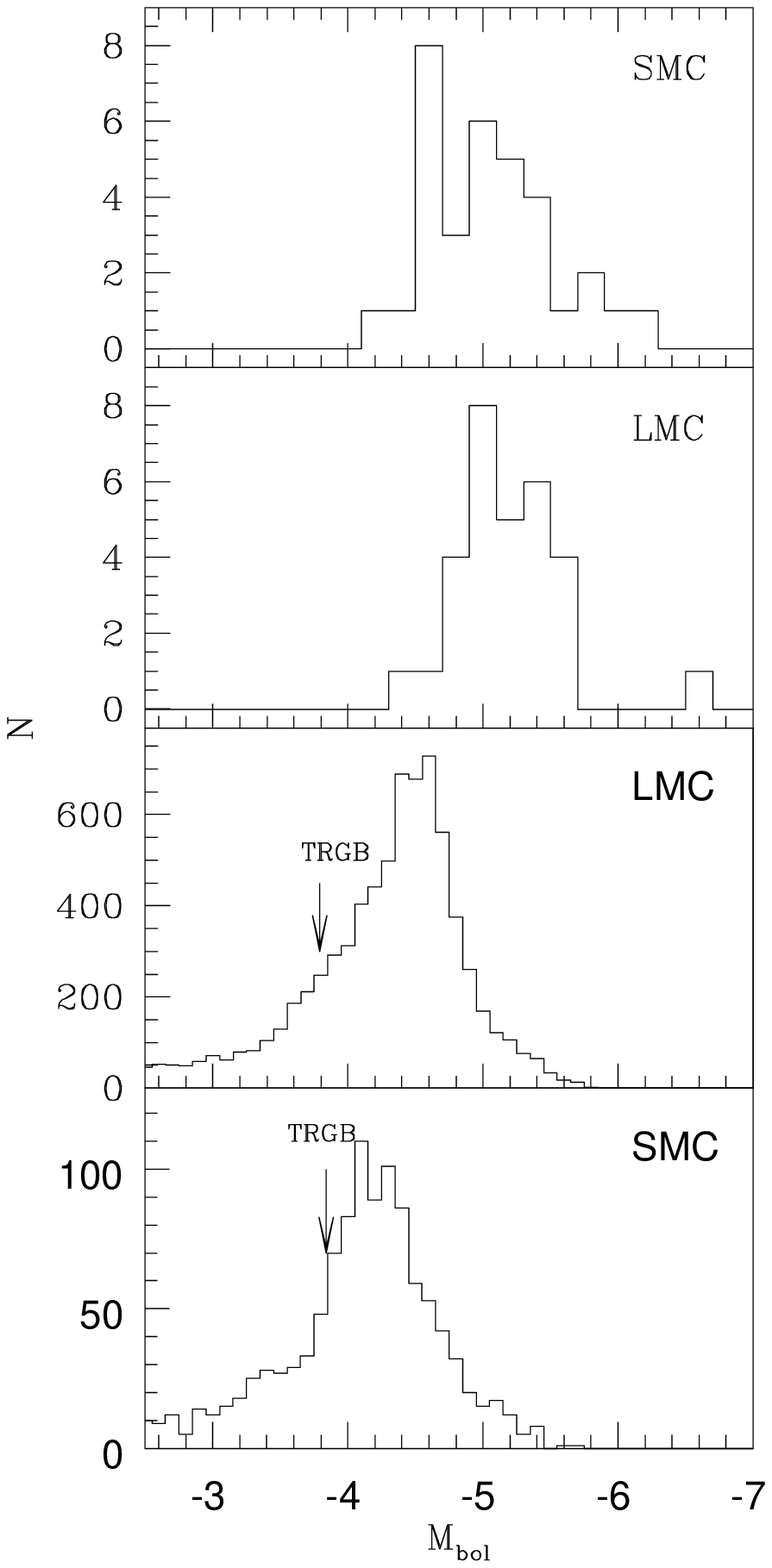}
\caption{\label{histo.ps}Top: distribution of bolometric magnitudes of the 
observed stars in the SMC, assuming a distance modulus of 18.93 (Keller \&
Wood 2006). Both stars from this paper and from Sloan et al. (2006) are
included.  Second: the histogram for the LMC carbon stars observed by Spitzer
(Zijlstra et al. 2006). Third: Optical carbon stars in the LMC (Kontizas et
al. 2001).  Bottom panel: Optical carbon stars in the SMC (Rebeirot et
al. 1993).  The arrows indicate the location of the tip of the RGB (Bellazzini
et al. 2001).}
\end{figure}

The histogram of the distribution is shown in Fig. \ref{histo.ps}.  The SMC
sample includes the carbon stars observed by Sloan et al. (2006) which cover a
similar luminosity range.  The LMC carbon-star sample of Zijlstra et
al. (2006) is also shown.  There is no obvious difference between the LMC and
SMC distributions.

For comparison, in the third panel we show a complete carbon star luminosity
function for the LMC. This is derived from the carbon star survey of Kontizas
et al. (2001), which we cross-correlated with 2MASS. The bolometric magnitudes
for this sample are derived from the 2MASS JHK magnitudes, using the
bolometric correction equation derived by Whitelock et al. (2006):
\begin{eqnarray}
\nonumber {\rm BC_K} & = & +\, 0.972 + 2.9292\times(J-K)
  -1.1144\times(J-K)^2 \\
 & & +0.1595\times(J-K)^3 -9.5689\,10^3(J-K)^4
\end{eqnarray}
\noindent 
This relation differs by 0.2--0.3 mag from the one used by Costa \&\ Frogel
(1996) for $1< \rm J-K<2.5$ .  Positional agreement between the 2MASS object
and the carbon star is required to be better than 1 arcsec. Stars with
non-detections in one or more infrared band are not included.

The bottom panel shows the same for the SMC carbon stars taken from
Rebeirot et al. (1993). These stars were also cross-correlated with
2MASS, and  the bolometric magnitudes  derived from the
2MASS JHK magnitudes. The coordinates in Rebeirot et al. are less
accurate and we accepted co-identifications out to 2 arcsec. A larger
number of chance superpositions may be expected.

If we compare this histogram with Fig.\,20 in Vassiliadis \& Wood (1993), for
SMC metallicities, this indicates that the stars we observed have initial
masses in the range 1--4 M$_{\odot}$.

Six of the stars we observed are located in the cluster NGC 419.  This cluster
has an age of 1.2 $\times$10$^9$ yr, and a metallicity [Fe/H]=$-$0.60$\pm$0.21
(de Freitas Pacheco et al. 1998). The isochrones of Pietrinferni et al. (2006)
for this age and for $Z=0.004$, give an initial mass of the thermal-pulsing
AGB stars in this cluster of 1.82\,M$_\odot$ for standard models and
2.08\,M$_\odot$ for models using overshoot. (The difference shows the
considerable uncertainty in deriving stellar ages.) But their bolometric
magnitudes (see also van Loon et al. 2005) are associated with higher-mass
progenitors on the Vassiliadis \&\ Wood (1993) tracks. Thus the stars reach a
higher luminosity than predicted. This may indicate that the mass loss is
weaker (or the onset of the superwind later) than assumed for the evolutionary
model calculations (Zijlstra 2004).

\begin{table}
\caption[]{\label{lit.dat} Bolometric magnitudes and periods from
the literature. B83: Bessell et al. (1983); W91: Westerlund et al. (1991);
T97: Tanab\'e et al. (1997); vL05: van Loon et al. 2005; 
W89: Whitelock et al (1989); LE88: Lloyd Evans et
al. (1988); C03: Cioni et al (2003).
The distance modulus to the SMC is taken as 18.93.
  }
\begin{flushleft}
\begin{tabular}{llllllllllllllll}
\hline
     & {$M_{\rm bol}$} & Period & ref.\\
Star & [mag] & [days] \\ 
\hline
NGC\,419 LE\,27 & $-4.88$, $-4.98$ &      & B83,W91\\
NGC\,419 LE\,18 & $-4.72$, $-4.82$ &      & B83,W91\\
NGC\,419 LE\,16 &  $-5.43$         &      & W91\\
NGC\,419 LE\,35 &  $-5.02$         &      & W91\\
NGC\,419 IR\,1  &         $-4.9$, $-5.3$    &      & T97, vL05\\
NGC\,419 MIR,1  &        $-4.9$    &      & vL05\\
IRAS\,00554$-$7351 &     $-6.1:$   &  800 & W89\\
LEGC\,105       &        $-4.7$    & 310:,350 & LE88,C03\\
RAW\,960        &                  & 34   & C03 \\
ISO\,00573      &        $-5.3 $   & 348  & C03 \\
ISO\,01019      &        $-5.1$    & 336  & C03 \\
ISO\,00548      &        $-4.9$    & 432  & C03 \\
ISO\,00549       &       $-5.6$    & 604  & C03 \\ 
\hline \\
\end{tabular}
\end{flushleft}
\end{table}

\begin{table*}
\caption[]{\label{targets.dat} Observed SMC targets:
names, adopted coordinates,  and  photometry.
J and K are taken from near-simultaneous measurements at SSO. The 
8$\mu$m and 24$\mu$m magnitude are calculated
from our spectra using the IRAC and MIPS filter profiles; the zeropoints are 
taken as 63.5 ( 8.0$\mu$m) and 7.14 Jy (24$\mu$m) respectively.
  }
\begin{flushleft}
\begin{tabular}{llllllllllllllll}
\hline
Adopted name &  2MASS name & RA & Dec  
& J & K & 8$\mu$m &24$\mu$m & m$_{\rm bol}$  & $M_{\rm bol}$  \\
  & &   \multicolumn{2}{c}{(J2000)}&mag  & mag & mag & mag & mag & mag\\
\hline
AGB stars\\
\\
NGC 419 LE 27 & 01082067$-$7252519           & 01 08 20.67& $-$72 52 52.0
&12.68 &10.99 &6.05 &\phantom{4.89} & 14.49 & $-4.44$ \\

NGC 419 LE 18& 01082495$-$7252569  & 01 08 24.95&$-$72 52 56.9 
& 12.71 & 11.00 & 9.09 &8.76& 14.43 & $-4.50$ \\

NGC 419 LE 35&  01081749$-$7253013          & 01 08 17.49& $-$72 53 01.3 
&12.67 &10.88 & 7.16 & \phantom{5.80} & 14.33  & $-4.60$ \\

NGC 419 LE 16&  01080114$-$7253173          &  01 08 01.14 &$-$72 53 17.4 
&13.82 & 11.30 &8.72 & & 14.65        & $-4.28$ \\

NGC 419 IR1&  01081296$-$7252439          &  01 08 12.97&$-$72 52 44.0
&13.44  & 10.74 & 7.31& 6.79& 13.71 & $-5.22$ \\

NGC 419 MIR1&             & 01 08 17.47& $-$72 53 09.5 
&       &15.49  &  & & 14.33  & $-4.60$ \\

IRAS 00554$-$7351&  00570395$-$7335146           & 00 57 03.95 & $-$73 35 14.7 
& 15.90 &11.09 & 5.69& 4.12& 12.67  & $-6.26$ \\

RAW 960   &      00555464$-$7311362        & 00 55 54.65 & $-$73 11 36.3 
& 13.29 & 10.92& & & 14.42          & $-4.51$ \\

ISO 00573&    00572054$-$7312460         & 00 57 20.55& $-$73 12 46.0
& 12.87 & 10.75 & 8.61&8.20 & 14.17 & $-4.76$ \\

LEGC 105&        00544685$-$7313376             & 00 54 46.85 & $-$73 13 37.7 
&14.23  &11.21 & 8.18 &7.85& 14.36 & $-4.57$ \\

ISO 01019    &    01015458$-$7258223            &       01 01 54.58    & $-$72 58 22.4
& 13.05 & 10.88 & 8.21 &8.12 & 14.15 & $-4.78$ \\

ISO 00548& 00545075$-$7306073 & 00 54 50.76	 & $-$73 06 07.4	
&13.87 & 10.87 &7.45 &7.12 & 13.84 & $-5.09$ \\

ISO 00549     & 00545410$-$7303181 	           &00 54 54.11  & $-$73 03 18.2
&14.59 &10.99 & 7.24&6.14 & 13.80 & $-5.13$ \\

GM780    &   00353726$-$7309561          & 00 35 37.27 & $-$73 09 56.2 
&12.88 & 10.18& 7.67& 6.57& 13.58  & $-5.35$ \\
\\
Other objects\\
\\

PMMR 52    &           & 00 53 09.12 & $-$73 04 03.8 
&  & & \\

MB 88 &               & 00 51 40.4 &$-$73 13 33\\
\hline \\
\end{tabular}
\end{flushleft}
\end{table*}


\begin{figure}
\includegraphics[width=9cm,clip=true]{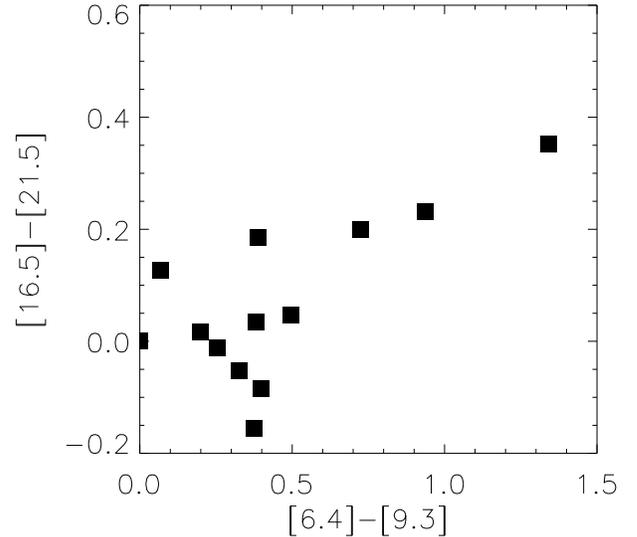}
\caption{\label{col_col.ps} The [6.4]$-$[9.3] versus [16.5]$-$[21.5] 
colour-colour diagram of the stars in our SMC sample.
 }
\end{figure}

\section{Observed spectral features}
 The spectra show clear and  deep molecular bands. The two main dust features
are less obvious. We will describe the dust bands first, followed by the
observed molecular bands. 

\subsection{Dust bands}
\label{dustbands}

Most of the carbon stars in our sample show the presence of an emission
feature around 11.3$\mu$m. This feature is attributed to emission from
SiC. However, in several cases the feature is very weak, and in such cases it
is hard to judge whether this feature is an emission feature or due to
molecular absorptions on the blue and red side. SiC condenses at high
temperatures and is expected to be present for all dusty carbon stars.
However, its abundance may be limited by the abundance of Si.

NGC419 LE 35 shows no discernable SiC feature. This is, however, one of the two
sources with acquisition and/or extraction problems. GM\,780 has a very strong
feature, compared to the other stars, but the shape is peculiar.
The shape is possibly affected by extraction problems but the
strength of the feature is  well established. This object
is discussed below.

A wide emission feature, centred at $\sim30\mu$m is observed in two stars: NGC
419 MIR1 and (weaker) IRAS 00554$-$7351. This feature is common in
Galactic AGB and post-AGB stars, and attributed to emission from MgS (Hony et
al. 2002).  These two SMC stars have the reddest near-infrared colours in our
sample, and the coolest dust, with dust temperatures of respectively 409 and
589 K (see section \ref{col_strength}).  In the LMC, this feature is observed
only in the envelopes of stars with dust temperature lower than 650K (Zijlstra
et al. 2006).  This can be explained by the fact that MgS grows on the surface
of pre-existing grains, this process starting around 600K and being complete
around 300K. The two SMC detections are consistent with this.

\subsection{Molecular bands}

Most vibrational bands of simple molecules occur at wavelengths shortward of
20$\mu$m. (HNC has a vibrational fundamental at 20$\mu$m.)  This is the region
where we observe clear bands.  The main features, observed for most of the
stars of our sample, are located at $\sim$7.5$\mu$m and $\sim$13.7$\mu$m
respectively. Both are associated with absorption from C$_2$H$_2$. The
13.7$\mu$m band is due to the $\nu_5$ bending mode of acetylene and its
associated hot and combination bands (Cernicharo et al. 1999).  This
absorption band is observed in almost all carbon stars of our sample, but it
is not clear in ISO\,00549 and GM\,780. The narrow band is flanked by broader
bands, best seen in NGC419 MIR1. The double peaked band at 7.5$\mu$m observed
in all of our spectra is associated with the C$_2$H$_2$ $\nu_4^1$+$\nu_5^1$ P-
and R-branch transition, as shown by Matsuura et al. (2006). The shape may be
different in NGC419 LE 27, perhaps with a contribution from CS and/or HCN.

A weak absorption band at 14.3$\mu$m is observed in the spectra of the two
reddest stars of our sample (IRAS 00554 and NGC 419 MIR1). This band has been
observed in Galactic carbon stars (Aoki et al. 1999) and is attributed to
HCN. The fact that it is observed in the reddest stars of our sample, where
the optical depth is important, and dust coolest, seems to indicate that this
molecule is present in the outer layers of the observed envelopes.

A decrease of the spectral energy distribution is observed near the blue edge
of all our spectra , around 5 $\mu$m. This is due to several molecules: C$_2$
and/or C$_3$ starting from 6$\mu$m (blueward), and CO absorption very close to
the 5 $\mu$m. Several targets show a higher flux level at the blue edge with a
sharp drop: this is not seen in the LMC spectra.

Zijlstra et al. (2006) have identified a weak absorption feature in the LMC
sample at 5.8$\mu$m, attributed to carbonyl. This feature is not detected in
our SMC sample. It could indicate an underabundance of CO with respect to
stars in the LMC, as expected given that there is less O to start with.

\subsection{Models of molecular opacity}

The top lines in Fig. \ref{spec_multi_offset_bands.ps} show the calculated
model spectra of a number of molecules. Similar plots in Zijlstra et al.
(2006) are based on the line lists and models of J{\o}rgensen et al. (2000).
Newer line lists are now available for some simple molecules (Harris et
al. 2002, 2006), such as HCN and HNC. We also revised opacity curves for CO,
C$_2$ and CS from Zijlstra et al. (2006).

The HCN/HNC linelist was computed using first principles quantum mechanics. It
contains about 160\,000 rotation-vibration energy levels truncated at 18000
cm$^{-1}$, and at $J=60$. There are around $4\times 10^8$ lines, but only
$34\times 10^6$ of these are strong enough to contribute to opacity.  Both the
linear H-CN and H-NC geometric configurations are studied, as are some of the
low lying delocalised states. The delocalised states are the states in which
the H nucleus orbits the CN part of the molecule. The accuracy of the
fundamental vibrational transitions is about 3--4 cm$^{-1}$; the error
increases at higher energies. The transition intensities agree well with
laboratory data. Incorporation of laboratory data into the linelist has
significantly improved the accuracy of the frequencies of the transitions
between these low lying states.

The opacity curves for CO, C$_2$ and CS are calculated from absorption
coefficients determined by Goorvitch (1994), Querci et al. (1994) and Chandra
et al. (1995) respectively.

Using this, we determined the opacity for the HCN, HNC, CO, CS and C$_2$ as a
function of wavelength. To determine the transmission function, we then used
the Lambert-Beer law (Banwell \&\ McCash 1994), i.e. the solution to the 1D 
equation of radiative  transfer in the absence of an internal source.

Fitting the spectral features to get column densities is beyond the scope of
this paper and will be done in a forthcoming paper. We can however determine
which molecules are responsible for the observed molecular bands. We thus
calculated transmission curves using column densities for the different
molecular species of $10^{18}$cm$^{-2}$ for the HCN, HNC, CS and C$_2$
molecules and $10^{22}$cm$^{-2}$ for CO. A temperature of 1750 K was used for
all the models. The resulting transmittance curves are plotted at the top of
Fig. \ref{spec_multi_offset_bands.ps}, together with the transmittance curves
from J{\o}rgensen et al. (2000) for C$_2$H$_2$ and C$_3$.  This C$_2$H$_2$
model reproduces the broad photospheric absorption, but not the narrow
component, arising from a cool layer (J{\o}rgensen et al. 2000).  Cherchneff
(2006) shows that acetylene reaches its peak abundance in the extended
atmosphere, and its temperature is expected to be lower than the photospheric
temperature (van Loon et al. 2006, Matsuura et al. 2006).
\begin{figure}
\includegraphics[width=9cm,height=6cm]{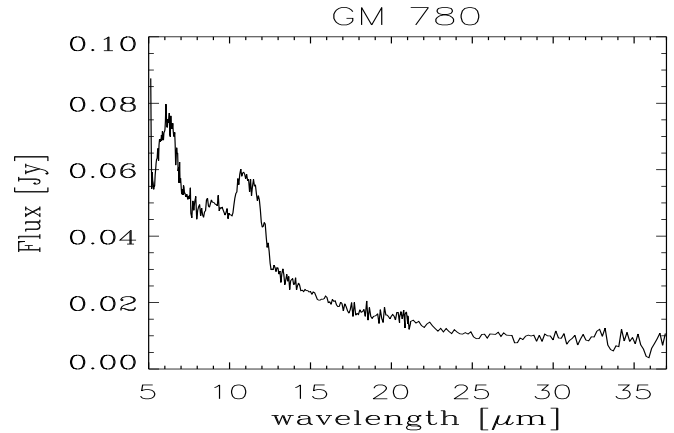}
\caption{\label{gm780.a.lo.ap.ps} The peculiar spectrum of GM\,780.}
\end{figure}

\section{Individual stars}

\subsection{PMMR 52}
\label{sectpmmr52}
PMMR\,52 is a red supergiant of spectral type K (Pr\'evot et al., 1983).  This
is the only oxygen-rich star in the sample. The spectrum is shown in
Fig. \ref{gm106.a.lo.ap.ps}.  The observed spectrum of PMMR\,52 shows the
presence of spectral features of oxygen-rich dust (silicates at 10$\mu$m and
18$\mu$m). The 18$\mu$m silicate feature is strong. The 10-$\mu$m silicate
band is remarkably weaker than observed in other O-rich stars. Normally, in
O-rich stars, the 18-$\mu$m feature should decrease steeply after
19-20$\mu$m. But the one in PMMR 52 remains very flat.

  There are no signatures of crystalline silicates, which indicates that the
silicates are largely in an amorphous phase. Some small absorption features
are observed at 6.2 and 7.5$\mu$m. The 6.2$\mu$m feature is due to water,
while the one at 7.5$\mu$m might be attributed to absorption by SiO, but this
feature is very faint and might be due to noise.

\subsection{GM 780}
\label{gm780}
Fig. \ref{gm780.a.lo.ap.ps} shows the spectrum of GM\,780. Compared to the
other sources, C$_2$H$_2$ bands are noticeably absent. The weak band at
8\,$\mu$m is best fitted as CS, based on the models above. The 11-$\mu$m band
is strong but its shape is significantly different from that of the other
stars.  There are no traces of silicates at 9.8 and 18$\mu$m. GM\,780 is the
brightest star at K in our sample; the bolometric luminosity is high for our
sample but lower than those of the luminous IRAS-selected carbon stars (van
Loon et al. 1999b).

 The $J-K$ colour of this object, $J-K=2.6$, is quite red, and dust should be
present around it. The dust temperature is relatively high, however, at
720\,K.

The object is discussed further below (Section~\ref{sic}) where we argue
the this star has a C/O ratio lower than that of other stars in the sample.

\section{Colours and band strengths} 
\label{col_strength}
\begin{center}
\begin{table}

\caption[]{\label{colours.def} Mid-infrared 
continuum bands for carbon stars for the so-called 'Manchester system'. 
The last column
gives the adopted flux corresponding to zero magnitude.
  }
\begin{flushleft}
\begin{tabular}{ccclll}
\hline
central $\lambda$ [$\mu$m] & $\lambda$-range [$\mu$m]  & $F_0$ [Jy] \\
\hline

   6.4 & 6.25--6.55 & 96.5 \\

   9.3 & 9.1--9.5   & 45.7 \\

   16.5 & 16--17    & 15.4 \\

   21.5 & 21--22   & 9.1 \\
\hline \\
\end{tabular}
\end{flushleft}

\end{table}
\end{center}


\begin{table}
\caption[]{\label{colors.dat} Photometry: colours measured using
four narrow carbon stars continuum bands. 
  }
\begin{flushleft}
\begin{tabular}{lcrllllllllllll}
\hline
target         & [6.4]$-$[9.3] & [16.5]$-$[21.5] \\
           
& [mag] & [mag] \\

\hline
NGC 419 LE 27     & 0.067 $\pm$0.018& \phantom{0.429$\pm$0.009} \\
NGC 419 LE 18     & 0.067 $\pm$0.020& 0.126$\pm$0.098 \\
NGC 419 LE 35     & 0.215 $\pm$0.013& \phantom{0.476$\pm$0.028} \\
NGC 419 LE 16     & 0.325 $\pm$0.021&$-$0.052$\pm$0.062 \\
NGC 419 IR1       & 0.496 $\pm$0.011& 0.047$\pm$0.017 \\
NGC 419 MIR1      & 1.343 $\pm$0.012& 0.352$\pm$0.012 \\
IRAS 00554        & 0.938 $\pm$0.008& 0.232$\pm$0.021 \\
RAW 960           & 0.199 $\pm$0.022& 0.017$\pm$0.115 \\
ISO 00573         & 0.255 $\pm$0.017&$-$0.012$\pm$0.029 \\
LEGC 105          & 0.380 $\pm$0.013& 0.034$\pm$0.069 \\
ISO 01019         & 0.374 $\pm$0.015&$-$0.156$\pm$0.039 \\
ISO 00548         & 0.399 $\pm$0.013&$-$0.084$\pm$0.025 \\
ISO 00549         & 0.725 $\pm$0.014& 0.199$\pm$0.012 \\
GM 780            & 0.388 $\pm$0.016& 0.185$\pm$0.031 \\

\hline \\
\end{tabular}
\end{flushleft}
\end{table}

As discussed above, the spectra of the observed stars are dominated by
molecular and dust bands. This makes the definition of the continua
difficult.  To avoid this and to determine colours, we use the Manchester
system (Zijlstra et al. 2006, Sloan et al. 2006).  This system, valid for
carbon stars but not for oxygen-rich stars, defines four narrow bands, selected
to avoid the molecular and dust bands, to determine the continuum.
Table~\ref{colours.def} shows the wavelengths used to defined the four
narrow-bands and the adopted flux corresponding to zero-magnitude, determined
to give a zero colour for a Rayleigh-Jeans tail.  Using this system, we can
determine continuum flux at 6.4, 9.3, 16.5 and 21.5$\mu$m, by integrating the
observed spectra over the defined bands.  The choice of the wavelengths used
to define the continua is discussed in Zijlstra et al. (2006).
Table~\ref{colors.dat} lists the measured continuum flux and colours for the
observed carbon stars.

This method permits us to determine two colour temperatures, [6.4]$-$[9.3] and
[16.5]$-$[21.5].  The [6.4]$-$[9.3] colour is an indication of the optical
depth, whereas the [16.5]$-$[21.5] colour can be use to determine the dust
temperature.  Table~\ref{ew.dat} lists the blackbody temperatures derived from
the [16.5]$-$[21.5] colours for the observed stars.  For the bluest objects,
this method gives lower limits ($\lsim 10^4$K) as the Rayleigh-Jeans limit is
reached.  This is the case for NGC\,419\,LE\,16, NGC\,419\,IR1, RAW\,960,
LEGC\,105, ISO\,00573, ISO\,01019 and ISO\,00548.  We thus do not give an
estimation of the dust temperature for these stars. Furthermore, the derived
temperature can be affected by wavelength-dependent continuum contamination
from other sources in the beam. Thus, for the bluest stars, we can obtain
temperatures that are not physically meaningful.

Fig. \ref{col_col.ps} shows the [6.4]$-$[9.3] vs [16.5]$-$[21.5] colour-colour
diagram obtained with the Manchester method.  The stars from our sample form a
well-defined sequence on this diagram.  Some stars show a negative
[16.5]$-$[21.5] colour (NGC\,419\,LE\,16, ISO\,00573, ISO\,01019, ISO\,00548).
These stars are very blue and their spectra are typical of naked carbon stars.
Two blue stars, NGC\,419\,LE\,18 and GM\,780, are offset in [16.5]$-$[21.5]
colour by roughly 0.1 mag.  The spectrum of GM\,780 is very noisy in LL, so
that its [16.5]$-$[21.5] colour has higher uncertainties than the other
objects.  The good correlation observed between [6.4]$-$[9.3] and
[16.5]$-$[21.5] indicates that the choice of our continuum wavelengths is
appropriate.

For all the stars in our sample, we have measured the strength of the observed
features.  We define small wavelength ranges on the blue and red side of the
features (Table \ref{cont.def}), selected to avoid features, to define the
continuum.  We then fit line segments from both sides of the feature that
defines the continuum.  As the red edge of the MgS feature is outside of the
IRS wavelength range, this method can not be applied to the MgS feature.  The
continuum under the MgS feature is thus assumed to be a blackbody with a
temperature deduced from the [16.5]$-$[21.5] colour.  After subtraction of the
continuum, we can measure the strength of each feature.  For the molecular
bands (C$_2$H$_2$ at 7.5 and 13.7$\mu$m), this strength is estimated as an
equivalent width.  For the dust emission features (SiC and MgS), we use a
line-to-continuum ratio, defined as the integrated flux of the band divided
by the integrated underlying continuum, over the wavelength range of the
feature.  The MgS feature extend beyond the edge of the spectral coverage:
its strength relates only to the part blueward of 38$\mu$m, and the continuum
is calculated as a black body (see Zijlstra et al. 2006). 

We also measured the central wavelength of the SiC band.  This central
wavelength is defined as the wavelength at which, after continuum removal,
the flux on the blue side equals the flux on the red side.  Table
\ref{ew.dat} lists the measured strength and central wavelengths of the
observed features. The MgS feature central wavelength can not be measured 
from the available wavelength range.

\begin{table}
\begin{center}
\caption[]{\label{cont.def} Wavelengths used to estimate the continua for the
  SiC and C$_2$H$_2$ spectral features.}
\begin{flushleft}
\begin{tabular}{clll}
\hline
Features & $\lambda$ [$\mu$m] & Blue continuum [$\mu$m]& 
Red continuum [$\mu$m]\\
\hline
C$_2$H$_2$ &    \phantom{0}7.5  &  \phantom{0}6.08--6.77    &  
 \phantom{0}8.22--8.55 \\
SiC        &   11.3 & \phantom{0}9.50--10.10   & 12.80--13.40 \\
C$_2$H$_2$ &   13.7 & 12.80--13.40  & 14.10--14.70 \\

\hline \\
\end{tabular}
\end{flushleft}
\end{center}
\end{table}

\begin{table*}
\caption[]{\label{ew.dat} Strength of the molecular and dust features, in
terms of either the equivalent width in microns, or the integrated
line-to-continuum ratio (Sect. \ref{col_strength}. The last column gives the
continuum (black-body) temperature, derived from the [16.5]$-$[21.5] colour
listed in Table \ref{colors.dat}}
\begin{flushleft}
\begin{tabular}{lrrrrrrlllllllll}
\hline
target         &  EW (7.5 $\mu$m)  & EW (13.7 $\mu$m) & L/C(SiC) &$\lambda_c$    & L/C (MgS)& T(K) \\

\hline
NGC 419 LE 27     &0.080 $\pm$0.015 &0.057 $\pm$0.009 &0.033$\pm$0.009  &11.96
$\pm$0.17 & & \phantom{346 $\pm$6} \\
NGC 419 LE 18     &0.101 $\pm$0.009 & 0.151 $\pm$0.038 &0.032$\pm$0.016 &11.46 $\pm$0.20 &   &1038$\pm$430 \\
NGC 419 LE 35     &0.129 $\pm$0.006 &0.079 $\pm$0.006 &$-$0.009$\pm$0.004&
10.48 $\pm$0.00 & &  \phantom{317.$\pm$ 15}\\
NGC 419 LE 16     &0.158 $\pm$0.011 & 0.033 $\pm$0.009 &0.056$\pm$0.009 &11.33 $\pm$0.06 & &\\
NGC 419 IR1       &0.150 $\pm$0.002 &0.044 $\pm$0.005 & 0.038$\pm$0.004 &11.05 $\pm$0.07 & &\\
NGC 419 MIR1      &0.187 $\pm$0.011 & 0.066 $\pm$0.005 &0.101$\pm$0.005 &11.38 $\pm$0.05 & 0.392 $\pm$0.015& 409$\pm$ 12\\
IRAS 00554        &0.102 $\pm$0.008 &0.079 $\pm$0.004 & 0.119$\pm$0.008 &11.28 $\pm$0.08 & 0.276 $\pm$0.026 & 589$\pm$44\\
RAW 960           &0.230 $\pm$0.009 & 0.051 $\pm$0.010 &0.057$\pm$0.010 &11.20 $\pm$ 0.11&  &  \\
ISO 00573         &0.148 $\pm$0.005 & 0.078 $\pm$0.013 &0.056$\pm$0.011 &11.19 $\pm$0.18 &  &  \\
LEGC 105          &0.116 $\pm$0.003 & 0.062 $\pm$0.006 &0.019$\pm$0.004 &11.01 $\pm$0.11 &  &  \\
ISO 01019         &0.151 $\pm$0.005 & 0.032 $\pm$0.013 &0.033$\pm$0.009 &11.26 $\pm$0.13 &  &  \\
ISO 00548         &0.138 $\pm$0.003 & 0.068 $\pm$0.002 &0.068$\pm$0.004 &11.27 $\pm$0.04 &  & \\
ISO 00549         &0.047 $\pm$0.013 & $-$0.041$\pm$0.010 &0.061$\pm$0.012 &11.21 $\pm$0.30 &  & 674$\pm$34 \\
GM 780            &0.149 $\pm$0.010 & 0.006 $\pm$0.007 &0.289$\pm$0.011 &11.13 $\pm$0.05 &  & 724$\pm$ 94\\

\hline \\
\end{tabular}
\end{flushleft}
\end{table*}


\section{Colour-colour discriminatin  between C and O-rich AGB stars}

Classification criteria for separating C-rich from O-rich
stars proposed by Egan et al. (2001) are based on the K$-$A versus J$-$K 
colours (where A is the MSX band centred at 8.3$\mu$m). These are now known
not to work effectively (Zijlstra et al. 2006, Buchanan et al. 2006).
Better separation is possible using different MSX bands (Whitelock et al.
2006) but in the Magellanic Clouds only the A-band was sensitive enough to
detect the AGB stars.

A variety of colour-colour plots based on IRAC and MIPS observations, useful
in distinguishing chemical type, are discussed in Blum et al. (2006).  These,
however, rely on the IRAC bands outside of the IRS wavelength coverage.  We
investigated whether a colour-colour diagram, based on Spitzer IRAC and MIPS
colours, could discriminate O-rich and C-rich stars using $\lambda>5\,\mu$m.
The equivalent IRAC and MIPS flux for the observed objects were derived from
the IRS spectra and the transmission curves of the IRAC 5.8 and 8.0$\mu$m
filters and MIPS 24$\mu$m filter.  As our sample contains just one O-rich
star, we added IRS spectra of evolved stars in the SMC and in the LMC
from different programs.

\begin{figure}
\includegraphics[width=6cm,clip=true]{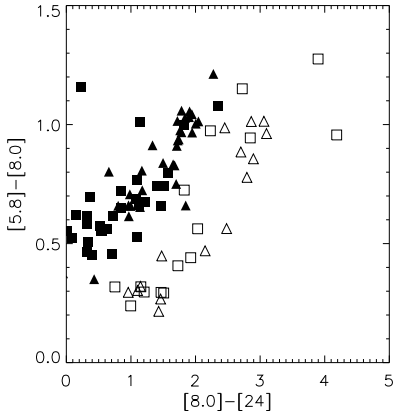}
\caption{\label{col_col_all} [8]$-$[24] vs [5.8]$-$[8.0] colour-colour 
diagram of AGB stars in the LMC and SMC. Black symbols represent AGB C-rich
stars. Open symbols represent AGB O-rich stars. Squares correspond to stars in
the SMC and triangles to stars in the LMC.}
\end{figure}

\begin{figure}
\includegraphics[width=6cm,clip=true]{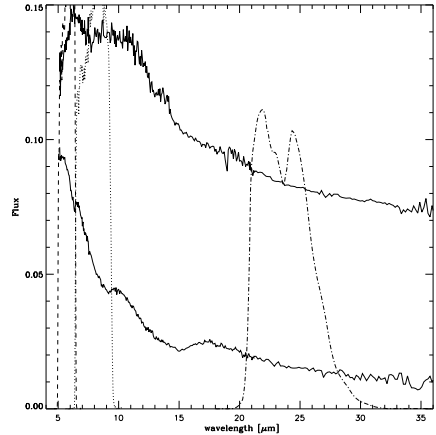}
\caption{\label{over} Spectra of two stars with the same [8]$-$[24] colours.
No flux scaling has been applied.
ISO\,00549 (top) is a C-rich star. PMMR\,52 (bottom) is an O-rich star. The
transmission of the IRAC 5.8$\mu$m (dashed line), 8$\mu$m (dotted line) and
MIPS 24$\mu$m (dot-dashed line) filters are overlayed. }
\end{figure}

Fig.\ref{col_col_all} shows the distribution of the O-rich stars (open
symbols) and C-rich stars (black symbols) from the combined sample on a
[8.0]$-$[24] vs [5.8]$-$[8.0] colour-colour diagram.  The O-rich and C-rich
stars form two well-defined sequences on this diagram.  The C-rich stars have
redder [5.8]$-$[24] colour than O-rich stars for the same [8.0]$-$[24] colour.
Fig.\ref{over} explains this effect.  We have plotted the spectra of PMMR 52,
an O-rich star, and ISO 00549, a C-rich star, with the transmission of the
IRAC 5.8um (dashed line), 8um (dotted line) and MIPS 24um (dotted-dashed line)
bands superposed.  The filter responses are affected by several molecular and
dust bands.  For O-rich stars, the blue edge of the MIPS 24$\mu$m filter
overlaps with the 18$\mu$m silicate emission feature, and molecular absorption
by SiO affects the 8$\mu$m band.  For carbon stars, the 5.8$\mu$m band is
suppressed by molecular absorptions. (We note that this IRAC band extends
beyond the blue edge of our spectra.)  A C-rich star with the same [8]$-$[24]
colour as an O-rich star will have a continuum emission redder than the O-rich
star. Such a diagram has been made by Buchanan et al. (2006) using their LMC
sample containing redder stars.

Fig.\ref{over} thus indicates that a [8]$-$[24] vs [5.8]$-$[8.0] diagram can
be used to discriminate between C and O-rich stars with colours within the
range of the objects present in our sample ([8]$-$[24] $<$4 and [5.8]$-$[8.0]
$<1.4$).  Furthermore, the sample used in Fig.\ref{col_col_all} contains stars
in the LMC (triangles) and the SMC (squares).  The SMC and LMC follow the same
sequences, indicating that the difference of location of C-rich and O-rich
stars in the colour--colour diagram is independent of metallicity.

\section{Average spectra for SMC and LMC stars}

\begin{figure*}
\includegraphics[width=17cm,clip=true]{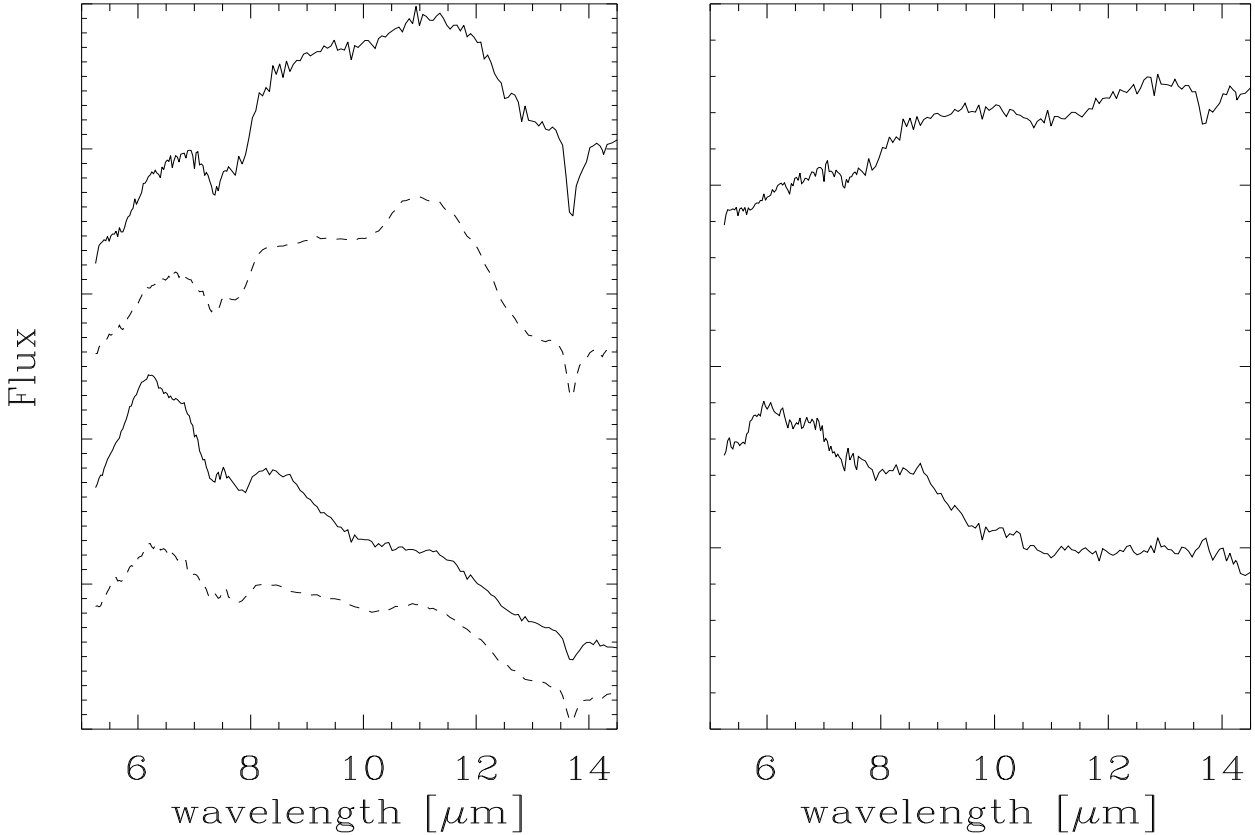}
\caption{\label{average} On the left, averaged spectra from our SMC sample 
(solid lines). The upper solid line shows the averaged spectrum of the red
sources. The lower solid line shows the averaged spectrum of the blue
sources. The dotted lines show the equivalent averaged spectra from the
Zijlstra et al. LMC sample. The right panel shows the SMC averaged 
spectrum divided by the LMC averaged  spectrum for the red sources (top) 
and the blue sources  (bottom).}
\end{figure*}

Fig. \ref{average} shows the averaged spectra of the  red SMC sources
(top, solid line) and blue SMC sources (bottom, solid line), where 'red'
and 'blue' are defined as in Zijlstra et al. (2006). To compare these
spectra with LMC spectra, we overplot the LMC spectra on this figure
(dotted lines). Those LMC spectra are taken from Zijlstra et al., and are
respectively the spectra with weak SiC and no MgS in Fig.19 in that paper. The
right panel shows the ratio of the SMC averaged spectra and the LMC averaged
spectra.

The LMC and the SMC spectra show the same features, but the differences are
significant especially for the 'red' sources. The differences are associated
with the C$_2$H$_2$ and the SiC bands.  The fact that the acetylene features
show as absorption in the ratio spectra indicates that these bands are
relatively {\it stronger} in the SMC. The 13.7$\mu$m is clearest seen,
indicative of the colder circumstellar molecules. The 11.3$\mu$m band also
shows in absorption, but as this is an emission feature, the negative
footprint here shows it to be {\it weaker} in the SMC.

The 'blue' sources do not show easily interpretable differences. The broad
10$\mu$m absorption may contain a blue contribution from a molecular band
(e.g. Zijlstra et al.  2006), in addition to SiC. The redward continuum may be
affected by the photospheric broad 14$\mu$m C$_2$H$_2$ band.

\section{Silicon-Carbide Dust}
\label{sic}

\subsection{Strength}

Fig.\ref{average} clearly indicates that the 11.3 $\mu$m SiC feature is
stronger in the LMC than in the SMC, even if this feature is rather weak in
blue sources.  The divided spectra do not show any difference in the shape of
the SiC feature in the SMC and the LMC, but only in the band strength. Thus,
whereas the emission strength varies, the emission shape is constant. The
composition of the SiC grains is likely the same for the two environments.

\begin{figure*}
\includegraphics[width=17cm,clip=true]{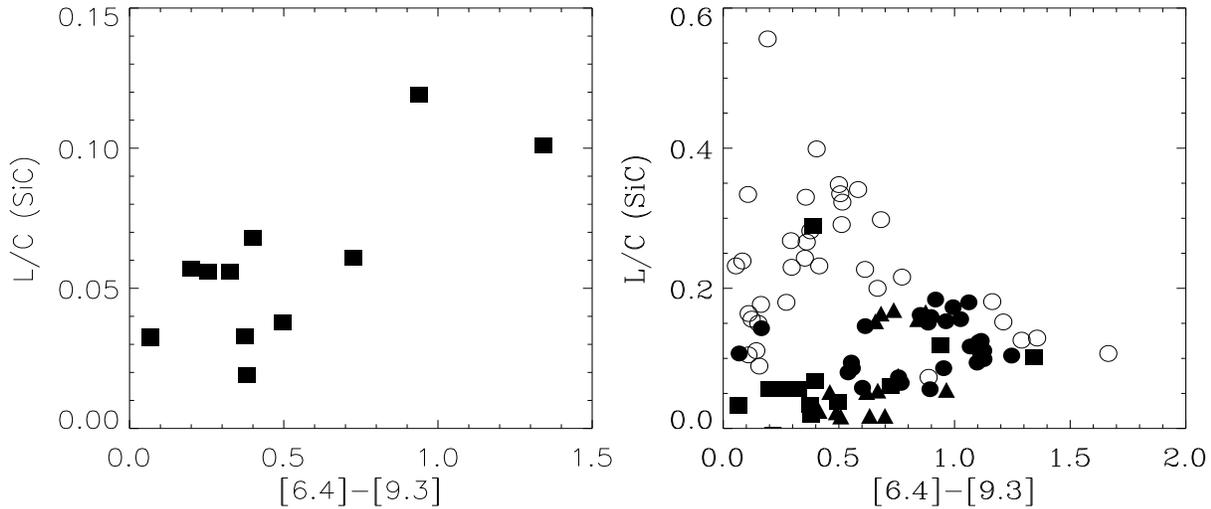}
\caption{\label{sicstr_col6.ps} The strength of the SiC feature as a function 
of the [6.4]$-$[9.3] colour in our SMC sample (left panel).  The right panel
combines this with stars from Sloan et al. SMC and Galactic samples: filled
squares represent our SMC sample, triangles Sloan et al. (SMC), open circles
Sloan et al. (Galactic) sample and filled circles Zijlstra et
al. (LMC).}
\end{figure*}

Fig.\ref{sicstr_col6.ps} shows the strength of the SiC feature as a function
of the [6.4]$-$[9.3] colour.  To study the effect of metallicity on the
strength of the SiC features, we overlaid the measured strength of this
feature on the Galactic and SMC sample of Sloan et al. (2006) as well as our
LMC sample (Zijlstra et al. 2006).  Sloan et al. (2006) find weaker SiC and
MgS features in the SMC than in the Galaxy, related to the lower abundance of
Si, S and Mg in the SMC.

Our SMC sample confirms that the strength of the SiC feature is lower in the
SMC than in the LMC and the Galaxy.  However, there are two noticeable
exceptions from the Sloan et al.\ sample (filled triangles).

The distribution of strength with colour (optical depth) is fundamentally
different in the Galaxy compared to the SMC.  In the SMC, there is a clear
trend of increasing SiC strength with optical depth (left panel of
Fig.\ref{sicstr_col6.ps}).  In the Galaxy, SiC increases rapidly up to
[6.4]$-$[9.3]$\approx0.5$, followed by a decline towards the SMC relation. The
difference in the Galactic and Magellanic Cloud distributions is very clear in
the right panel of Fig. \ref{sicstr_col6.ps}.  Note that one SMC star follows
the Galactic sample: GM\,780, located outside of the scale of the left panel
(and therefore not shown). This object (see Section \ref{gm780}) has a strong
SiC feature but with an unusual shape and rather weak dust continuum. We
return to it below.

The rapid increase of SiC L/C for Galactic stars at very low [6.4]$-$[9.3]
colour (Fig. \ref{sicstr_col6.ps}) suggests that in the Galaxy, the SiC
feature is present while the dust excess is still low. The subsequent decline
in L/C ratio may include effects of optical depth within the feature (Speck et
al. 2005), or it may be due to the increasing dust continuum.  The LMC stars
increase much more slowly with optical depth, and reach the peak around
[6.4]$-$[9.3]$\approx1.0$; redward they are close to or a little below the
Galactic stars. For the SMC stars, there is no well-defined peak but most SMC
stars with [6.4]$-$[9.3]$<1.0$ have little or no SiC.

\subsection{Condensation sequence}

The difference between the SMC and the Galaxy can be understood if for
Galactic stars, SiC condenses first (i.e. at higher temperatures) and carbon
dust later. For the LMC stars, the two form close together, while for the SMC
stars, SiC forms last.  Thus, there is a reversal in condensation sequence over
this range in metallicity.

The condensation temperature of graphite is around 1600\,K whilst for SiC $T_c
\approx 1400\,$K. However, these are composition and density dependent.  For
high partial pressure, condensation occurs at higher temperatures. The effect
is shown by Bernatowicz et al. (2005). For increasing C/O ratio, graphite
condenses at higher $T$. With decreasing Si abundance, $T_c({\rm SiC})$ will
decrease. The effect is that SiC condenses first at high density, low C/O
ratio ($<1.1$) and high Si abundance. Fig. \ref{sicstr_col6.ps} suggests these
conditions are met for Galactic stars. For LMC stars, SiC condensation is
delayed, while for SMC stars in many cases SiC does not condense at all. The
increasing C/O ratio also aids this process.  Chigai \&\ Yamamoto (1993) also
suggest that under different conditions,  SiC and graphite may form
together, only SiC grains may form or only graphite forms.

Following this, we interpret GM\,780 as a case where the density in the
dust-formation region is high and the C/O ratio low, leading to SiC only.  The
relatively blue Manchester colour of this object is due to the resulting lack
of dust. The lack of clear C$_2$H$_2$ features is consistent with this
interpretation.

If the condensation sequences discussed above are correct, the SiC abundance
may be limited by Si in the Galaxy (where SiC condenses first) but by the C/O
ratio in the Magellanic Clouds (where much of the carbon goes into the dust
before SiC can form).

\subsection{Central wavelength}

\begin{figure*}
\includegraphics[width=17cm,clip=true]{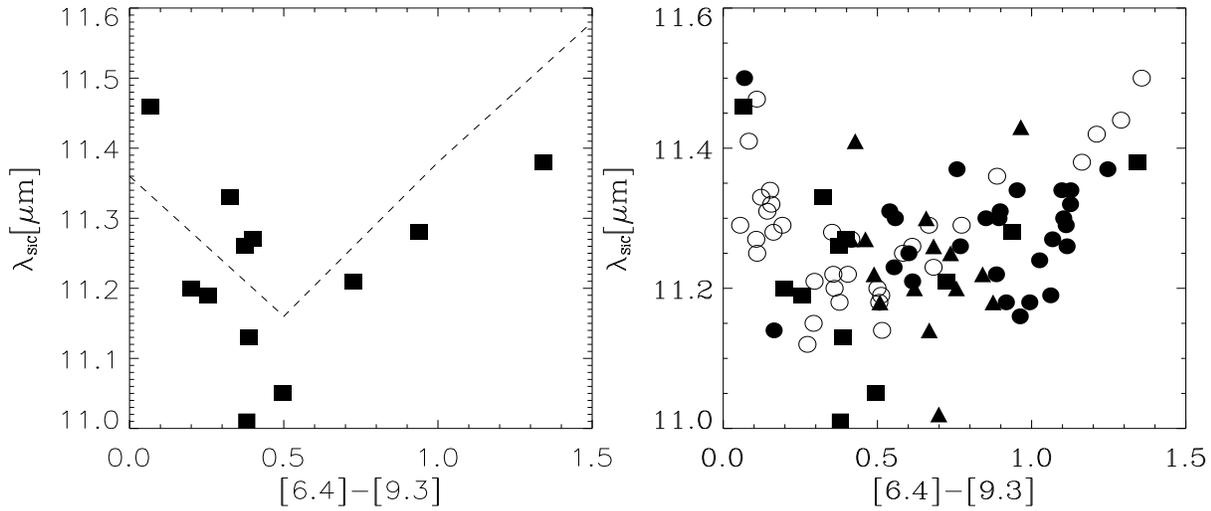}
\caption{\label{col_lambdasic.ps} Apparent central wavelength of the SiC 
feature as a function of the [6.4]$-$[9.3] colour. Left panel: our SMC
 sample. Right panel: filled squares represent our SMC sample, triangles
 Sloan et al. (SMC), open circles Sloan et al. (Galactic) sample and filled
 circles Zijlstra et al. (LMC).  The dashed line is the fit for Galactic stars
(Sloan et al. 2006).}
\end{figure*}

Fig. \ref{col_lambdasic.ps} shows the apparent central wavelength of the SiC
feature as a function of the [6.4]$-$[9.3] colour.  The right panel displays
our measurements together with the other samples mentioned above.  Our sample
shows that for the stars with [6.4]$-$[9.3]$<$0.5, there is a clear trend of
decreasing central wavelength with increasing colour, while the opposite is
observed for redder stars.  The dashed line shows the fit proposed for
Galactic stars (Sloan et al. 2006).

This shift could be due to two factors (Zijlstra et
al. 2006). The broad C$_2$H$_2$ absorption band centred at 13.7$\mu$m is close
to the SiC feature and the red continuum wavelengths we use for SiC necessarily
fall within this band. Thus stronger C$_2$H$_2$ absorption shifts the SiC
central wavelength to the red. Speck et al. (2005) also suggested that
self-absorption of SiC can also induce a shift toward the red of the SiC
feature.

The shape of the SiC feature appears to be the same for all stars
(e.g. Zijlstra et al. 2006). This argues against wavelength-dependent optical
depth effects (Speck et al. 2005) for our samples. We note that the stars
studied here have relatively low optical depth compared to stars with known
SiC absorption.

\section{Magnesium sulfide}

Longward of 15$\mu$m, only  the MgS feature is seen. The
shape of this feature is similar in the SMC and the LMC. Furthermore, we have
shown (Sect.\ref{dustbands}) that this feature is present in the envelopes
of the reddest stars, i.e. the stars with the lowest dust temperature. This
is explained by the formation process of MgS which starts around 600K and is
complete around 300K.

\begin{figure}
\includegraphics[width=9cm,clip=true]{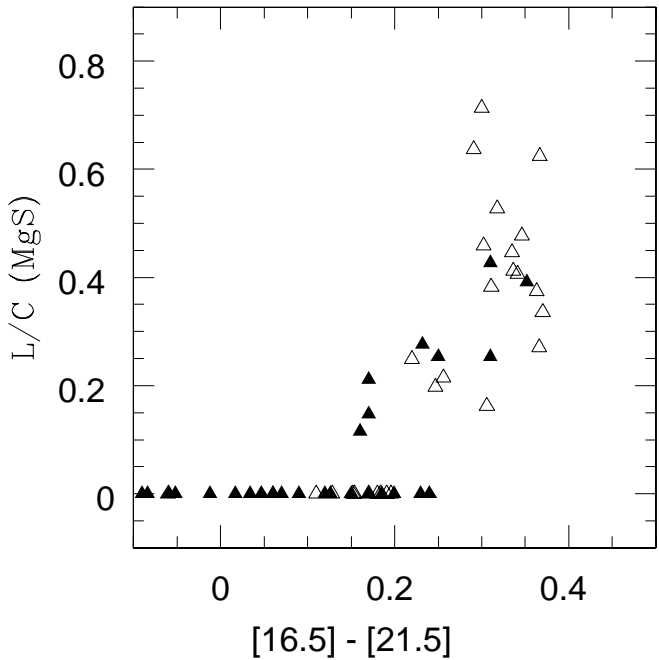}
\caption{\label{mgs.ps} Strength of the MgS feature,
versus [16.5]$-$[21.5] colour.  Filled triangles: SMC stars (present sample
and Sloan et al. 2006); open triangles: LMC stars (Zijlstra et al. 2006). 
The band strength is defined as the integrated line-to-continuum ratio (Sect.
\ref{col_strength}. }
\end{figure}

The feature is only seen for two stars in the SMC sample, but this appears to
be due to the dust temperatures. Fig. \ref{mgs.ps} shows the comparison
between the SMC and the LMC. There is little difference between them,
and it appears the ratio between MgS and carbon dust is less sensitive to
the metallicity than is the SiC versus carbon dust. This may indicate that
the MgS condensation is limited by the available dust surface rather than
the element abundances.

\begin{figure}
\includegraphics[width=9cm,clip=true]{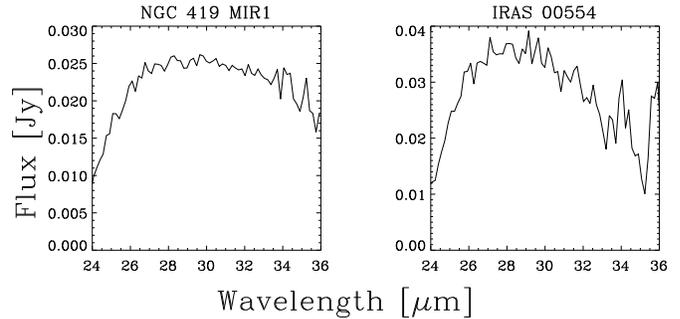}
\caption{\label{mgsfeat}Shape of the MgS feature, continuum subtracted, 
for SMC carbon-rich AGB stars.}
\end{figure}

The shape of the continuum subtracted MgS feature is shown in
Fig.\ref{mgsfeat}.  The MgS band of NGC 419 MIR1 has a relatively flat shape,
while the band observed in IRAS 00554$-$7351 diminishes toward the red. These
shapes are similar to the shapes of the MgS features observed in the LMC
(Zijlstra et al. 2006). The study of this feature in Galactic sources has
shown that it could be resolved in two subpeaks centred at 26 and 33$\mu$m
(Volk 2002). Those subpeaks are not observed in our sample, but this is not
conclusive due to the low signal.

\section{Gas properties}

\subsection{C$_2$H$_2$}

\begin{figure*}
\includegraphics[width=17cm,clip=true]{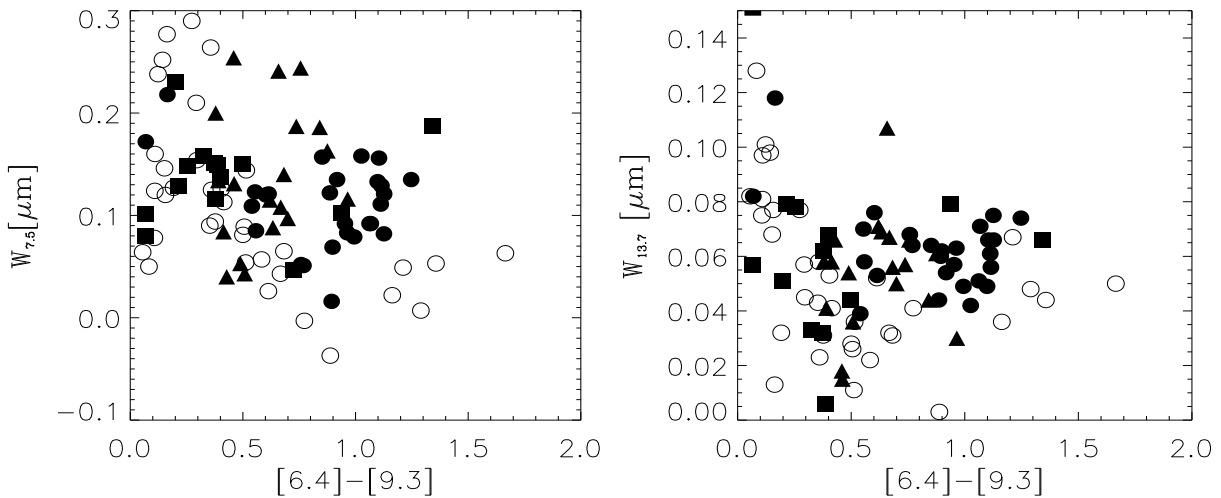}\hspace{-2cm}
\caption{\label{c2h2.eps} The equivalent width of the C$_2$H$_2$ band 
at 7.5$\mu$m (left) and 13.7$\mu$m (right) as a function of the [6.4]$-$[9.3]
colour. Filled squares represents our SMC sample, triangles the Sloan et
al. SMC sample, open circles the Sloan et al.  Galactic sample and filled
circles the Zijlstra et al. LMC sample.  }
\end{figure*}

Fig. \ref{c2h2.eps} shows the equivalent width of the C$_2$H$_2$ band at
7.5$\mu$m, for SMC, LMC and Galactic stars.  Due to a lower dust-to-gas ratio
at lower metallicity, stars with the same gas mass-loss rate will have a lower
optical depth in the SMC than in the LMC and the Galaxy.  If we use the
[6.4]$-$[9.3] colour as a tracer of the {\it dust} mass-loss rate, the
strength of the 7.5$\mu$m band is comparable in the Galactic sample and our
SMC sample. However, the [6.4]$-$[9.3] colour will be different for a given
{\it gas} mass-loss rate.  Zijlstra et al. (2006) show that the effect of
metallicity on the optical depth (lower metallicity stars have lower optical
depth) is stronger for redder stars (see Fig. 10 in their paper). Most of the
SMC stars in our sample have [6.4]$-$[9.3]$<$0.4.  To be compared with stars
with similar gas mass-loss rates, we should, according to the previous work,
compare those stars with stars with [6.4]$-$[9.3]$\approx0.5$ and 0.6 in the
LMC and the Galaxy, respectively.

By applying this shift, we observe that the 7.5$\mu$m feature is stronger in
the SMC than in the LMC and the Galaxy.  This feature appears to be stronger
in lower metallicity environments.  This has previously been found for the LMC
(Zijlstra et al. 2006, Matsuura et al. 2006) and the SMC (Sloan et al. 2006).
The stars in our sample are bluer than the previous observed stars, and we
confirm that the C$_2$H$_2$ feature is stronger in low metallicity
environments even when the stars are almost naked stars.

As noticed before in the LMC and SMC, this effect is more clearly seen for the
7.5$\mu$m feature than for the 13.7$\mu$m one.

The divided spectra in Fig. \ref{average} present absorption bands at 7.4 and
13.7$\mu$m. This confirms that the C$_2$H$_2$ features are stronger in the SMC
than in the LMC.  

\begin{figure}
\includegraphics[width=9cm,clip=true]{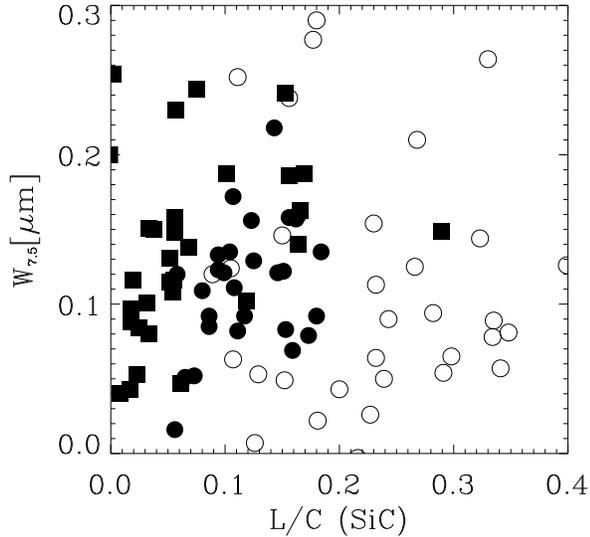}
\caption{\label{corsic} The strength of the C$_2$H$_2$ 7.5$\mu$m feature as a 
function of the SiC feature. Squares represent SMC stars from our sample and
Sloan et al. sample, black circles Zijlstra et al. LMC sample and open circles
Sloan et al. Galactic sample.}
\end{figure}

Fig.\ref{corsic} shows a trend of increasing C$_2$H$_2$ strength with
increasing SiC strength in the SMC. Such a trend is also observed in the LMC
but the slope is different. Whether this trend is present for the Galaxy is
less obvious. The SiC condensation sequence discussed above is consistent with
this: in the Galaxy, SiC is limited by Si, and in the Magellanic Clouds by the
C/O ratio. This predicts a correlation with C$_2$H$_2$ only for the Magellanic
Clouds.

The distribution in Fig.\ref{corsic} clearly shows a separation into
SMC--LMC--Galaxy, from left to right. The ratio of C$_2$H$_2$ (gas) over SiC
(dust) is higher in the SMC, then in the LMC and the Galaxy. This shows
an increasing gas-to-SiC dust mass ratio with decreasing metallicity.

%


\subsection{CO}
In all the spectra, the flux drops sharply shortward of 6$\mu$m. This is due
to absorption by the CO molecule (J{\o}rgensen 2000), C$_2$ and/or C$_3$
(Zijlstra et al. 2006), as shown by the spectra at the top of Fig.
\ref{spec_multi_offset_bands.ps}.  The slope of this feature is steeper in the
LMC than in the SMC.  We attribute this to an underabundance of CO in the SMC
with respect to the LMC. For carbon stars, the CO abundance is limited by
oxygen. Oxygen is a measure for the original metallicity: the CO band is
therefore expected to vary with metallicity.
Note that the responsivity at the blue edge of the spectra is low.

\section{Carbon versus Oxygen-rich stars}

A noticeable finding for both the LMC and the SMC is that the Spitzer targets
are found to be strongly biassed towards carbon stars. The selection criteria
were relatively insensitive to the chemical type, and the bias is therefore
intrinsic rather than a selection effect. Apart from stars observed as a
result of acquisition errors, we found 14 carbon stars and no oxygen-rich
stars.

Sloan et al. (2006) use 2MASS colours as part of their selection
criteria. This gave a sample of 22 carbon stars versus 10 oxygen-rich stars
(excluding some other sources). The use of 2MASS requires low circumstellar
reddening and favours stars with silicate dust where the reddening per unit
dust mass is lower.

The LMC sample of Zijlstra et al. (2006) shows 28 carbon stars versus one
oxygen-rich star (excluding one high-mass proto-star).  Buchanan et al. (2006)
observed high-luminosity stars and found 16 C-rich versus 4 O-rich AGB
stars. Thus, the dominance of carbon stars is confirmed in all surveys of
mass-losing stars, but the ratio depends on the characteristics of the survey.
The proportion of C stars is lower in high luminosity, lower-optical depth
surveys.

Cioni \&\ Habing (2003) show that for optically visible evolved stars in the
Magellanic Clouds, the ratio of C over M stars is approximately 3 for both
clouds.  Blum et al. (2006) surveying the full population of AGB stars in the
LMC, identify 17500 O-rich and 7000 C-rich stars. The higher ratio of O to C
stars found by Blum et al. may suggest that they include K-type stars, earlier
on the AGB. Cioni \&\ Habing (2003) estimate that there are about 2250 C stars
in the SMC and 10000 in the LMC, the latter being roughly consistent with the
Blum et al. numbers.

The high ratio ($\sim$3) of C to O-rich stars is consistent with the
expectation that in the Magellanic Clouds, most stars become carbon stars early
on the AGB, before the onset of the high mass-loss rate phase.  There are
almost no O-rich stars in the present and the Zijlstra et al. (2006) samples of
lower luminosity, high mass loss rate, AGB stars in the Magellanic Clouds.
Only among the highest mass, higher luminosity stars on the AGB do O-rich stars
contribute significantly to the population of mass-losing stars."

\subsection{Carbon star evolution}

Fig. \ref{histo.ps} shows that in the LMC, the carbon star luminosity function
exhibits a flat peak with a width of 0.3\,mag. This traces the part of the AGB
evolution where essentially all stars have become carbon stars: the flat
distribution results from the linear change of magnitude with time. (Stars
more massive than $\sim 4\,\rm M_\odot$ may avoid a carbon star phase due to
hot bottom burning, but there are few such stars.) At the high-luminosity end,
the number of known stars drops because of increasing
obscuration and  stars leaving the AGB.

In the SMC, the width of the flat peak is about 0.6\,mag.  Over $10^6\,$yr, a
star brightens by approximately 1\,mag on the AGB. These numbers therefore
predict that the optical carbon star phase lasts $3\times10^5$ to
$6\times10^5$\,yr, for the LMC and the SMC respectively. These are lower
limits to the total life time of the carbon star phase on the AGB.

The SMC carbon star distribution peaks at lower luminosity than in the
LMC. The longer life time in the SMC is therefore due to the stars becoming
carbon-rich earlier, after fewer thermal pulses.

The {\it Spitzer} targets all are at the high-luminosity end, and trace the final
stage of their evolution.  The Manchester colours of the SMC stars indicates
mass-loss rates of $10^{-5}\,\rm M_\odot\,yr^{-1}$ or higher (Zijlstra et
al. 2006, their Fig.  10).  Assuming that a typical carbon star expels
$0.2\,\rm M_\odot$ during the superwind, indicates a duration of this phase of
order $\sim 10^4\,\rm yr$. The combined SMC samples have approximately 30
superwind stars, or roughly 1 per cent of the total number of carbon stars
predicted by Cioni \&\ Habing (2003). This suggests a life time of the carbon
star phase of $\sim 10^6\,\rm yr$.  For comparison, the thermal-pulsing AGB
lasts 1--2$ \times 10^6$\,yr for 1--3\,M$_\odot$ stars. Thus, SMC stars become
carbon stars during the first half of the thermal-pulsing phase.

\subsection{Mass loss and metallicity}

A clear result from the {\it Spitzer} surveys is the enhancement of C$_2$H$_2$ in
low metallicity environments. This has previously been found from
ground-based spectra.  Van Loon et al. (1999b) and Matsuura et al. (2002,
2005) argue that this is related to a higher C/O ratio in these stars.  The
C/O ratio is enhanced by two effects: (i) a lower oxygen abundance means that
less carbon is locked up in CO (e.g. Lattanzio \&\ Wood 2004), and (ii) lower
metallicity promotes more efficient third dredge-up of carbon (Wood 1981,
Vassiliadis \&\ Wood 1993) with stars experiencing more thermal pulses (Lawlor
\&\ MacDonald 2006).  

This seems to be confirmed by the finding that LMC carbon stars have higher
abundances of C$_2$H$_2$ than Galactic stars.  In contrast, oxygen-rich stars
show low molecular abundances at low metallicity (Zijlstra 2006).

{\it Spitzer} shows that in the LMC and SMC, high mass-loss stars are
heavily dominated by carbon stars, unlike in the Galaxy (this paper, Zijlstra
et al. 2006, Sloan et al. 2006, van Loon et al. 2006, Buchanan et al. 2006,
Blum et al. 2006). In contrast, the sole oxygen-rich AGB star in the sample of
Zijlstra et al. has a rather moderate mass loss. Only the most luminous LMC
stars show an O-rich population with thick dust shells.

This leads to the suggestion that at low metallicity, stellar mass loss is
mostly carbon-rich, as at the luminosity where stellar pulsations develop an
extended atmosphere leading to molecule and dust formation, the stars have
already become carbon rich.  Only in old stellar systems (globular clusters,
Galactic halo), where the stellar masses are insufficient for third dredge-up,
is an oxygen-rich wind expected. 

The lack of silicon implies that the dust input into the ISM from LIMS in low
metallicity systems would be lacking in silicates and silicate-carbides, but
be dominated by carbonaceous dust, a 'soft' (mineral-poor) type of dust
(Zijlstra 2006, Matsuura et al. 2006, Groenewegen et al. 2007).

\section{Conclusion}

We have presented a {\it Spitzer} spectroscopic survey of 14 carbon-rich AGB stars in
the SMC. The bolometric magnitudes of the observed stars indicates that their
initial masses are in the range 1--4 M$_{\odot}$.  We also presented spectra of
an accidentally-observed K supergiant and an HII region.

Our spectra covers the range 5-38 $\mu$m. Molecular bands due to C$_2$H$_2$ at
7.5 and 13.7um are observed in most of the C-rich stars.  A weak absorption
band at 14.3 $\mu$m, attributed to HCN, is observed in the two reddest stars
of our sample. The 5.8-micron band of carbonyl observed in the LMC is not
observed in our SMC sample. This could be due to an underabundance of CO in
the SMC with respect to the LMC. The acetylene bands provide clear evidence
for a higher C/O ratio in low metallicity carbon stars.

Using the  ``Manchester System'', we determined the continuum flux
for the observed stars using four narrowbands centred at 6.4, 9.3, 16.5 and
21.5$\mu$m. This enables us to determine two colour temperatures,
[6.4]$-$[9.3] and [16.5]$-$[21.5]. These colours are indicators of the
optical depth and dust temperatures respectively.

We find evidence for a different behaviour of the SiC feature with
metallicity: in the Galaxy, it appears for stars with little dust excess,
while for the SMC it only appears for stars with cool dust. The LMC is
intermediate.  We suggest that this is due to a different formation sequence,
where at solar metallicity SiC precedes graphite, while at SMC metallicity and
its corresponding higher C/O ratio, graphite forms first.

The MgS feature at $\sim$30 $\mu$m has strength comparable to LMC stars,
suggesting its abundance relative to that of amorphous carbon is less
metallicity dependent. We suggest its formation is limited by the available
surface area of pre-existing dust grains, rather than elemental abundances.
The shape of the MgS and SiC features are similar in both galaxies, but the
SiC feature is much stronger in the LMC than in the SMC.

We show that a colour-colour diagram using Spitzer IRAC and MIPS
filters ([5.8]$-$[8.0] vs [8.0]$-$[24]) is able to separate O-rich and carbon
stars. This had been shown by Buchanan et al. 2006 for redder stars; we
confirm that it is also valid for blue stars ([8]$-$[24])$<$3).  Furthermore,
we show that this separation is independant of metallicity. This suggests
that it is possible to discriminate C and O-rich AGB stars in other galaxies 
using only {\it Spitzer} photometry.

Carbon stars dominate the population of high mass loss rate AGB stars in the
SMC. We derive a life time of the carbon star phase of $\gsim 6 \times
10^5$\,yr, with the superwind phase lasting $\sim 10^4$\,yr. Thus, stars
become carbon-rich early on the thermal-pulsing AGB, long before the onset of
the superwind. The LMC stars become carbon-rich later in their evolution, and
last as carbon stars for $\gsim 3 \times 10^5$\,yr.

\section*{Acknowledgments}
EL acknowledge support from a PPARC rolling grant. AAZ acknowledges a Royal
Society grant to allow a visit to the SAAO, and is grateful for the
hospitality of the SAAO. PRW has been partially supported in this research by
a Discovery Grant from the Australian Research Council.

\label{lastpage}

\end{document}